\pdfoutput=1

\documentclass[%
 reprint,
 amsmath,amssymb,
 aps,
]{revtex4-1}

\usepackage{graphicx}% Include figure files
\usepackage{dcolumn}% Align table columns on decimal point
\usepackage{bm}% bold math
\begin{document}

\preprint{APS/123-QED}

\title{Nucleotide-induced conformational motions and transmembrane\\gating dynamics in a bacterial ABC transporter}

\author{Holger Flechsig}
\email{holgerflechsig@hiroshima-u.ac.jp}
\affiliation{Department of Physical Chemistry, Fritz Haber Institute of the Max Planck Society,\\Faradayweg 4-6, 14195 Berlin, Germany\\}
\affiliation{Department of Mathematical and Life Sciences, Graduate School of Science, Hiroshima University, 1-3-1 Kagamiyama, Higashi-Hiroshima, Hiroshima 739-8526, Japan}

\date{\today}

\begin{abstract}
ATP-binding cassette (ABC) transporters are integral membrane proteins that mediate the exchange of diverse substrates across
membranes powered by ATP hydrolysis. We report results of coarse-grained dynamical simulations performed for the
bacterial heme transporter HmuUV. Based on the nucleotide-free structure, we have constructed a ligand-elastic-network
description for this protein and investigated ATP-induced conformational motions in structurally resolved computer experiments.
As we found, interactions with nucleotides resulted in generic motions which are functional and robust. Upon binding of ATP-mimicking
ligands the structure changed from a conformation in which the nucleotide-binding domains formed an open shape, to a conformation
in which they were found in tight contact and the transmembrane domains were rotated. The heme channel was broadened in the
ligand-bound complex and the gate to the cytoplasm, which was closed in the nucleotide-free conformation, was rendered open by a
mechanism that involved tilting motions of essential transmembrane helices. Based on our findings we propose that the HmuUV transporter
behaves like a `simple' mechanical device in which, induced by binding of ATP ligands, linear motions of the nucleotide-binding domains
are translated into rotational motions and internal tilting dynamics of the transmembrane domains that control gating inside the heme pathway.
\end{abstract}

\maketitle

\section*{Introduction}

ATP-binding cassette (ABC) transporters are molecular machines that are located in cellular membranes and involved in the active transport
of chemical substances \cite{higgins_92}. Owing to their enzymatic activity of binding ATP molecules and catalyzing their hydrolysis reaction,
they are capable of converting chemical energy into internal mechanical motions which are used to organize the transmembrane exchange of 
a variety of substrates such as ions, lipids, metabolic products, antibiotics and drugs \cite{rees_09}. In bacteria, ABC transporters play an
important role in the extrusion of toxins, and, on the other side, they can mediate the efficient uptake of nutrients \cite{fath_93,davidson_04,moussatova_08}.
ABC transporters are also clinically relevant since they can contribute to the resistance of cancer cells to drugs, to antibiotic tolerance, or cause
diseases such as cystic fibrosis \cite{gottesman_01,borst_02}. The common architecture of these transporters reveals two cytosolic nucleotide-binding
domains, the ATP-binding cassettes, and a pair of domains which are integrated into the membrane. Binding of intracellular ATP molecules and their
hydrolysis inside the cassettes is controlling the conformation of the transmembrane domains which contain the substrate-specific translocation
pathway, thus implementing the respective function of the transporter \cite{schneider_98,hollenstein_07,wen_11}.

The structures of several ABC transporters have become available in the past and it has been shown that the nucleotide-binding domains (NBDs)
are similar and constitute the engine core of the transporter, containing conserved sequence motifs which are involved in interactions with ATP
molecules  \cite{smith_02,chen_03,zaitseva_05,dawson_06,procko_06}. The transmembrane domains (TMDs), however, can considerably differ in their
sequence and architecture despite all of them having a central pathway for the translocation of substrates.
Moreover, since a few structures of transporters in the presence of nucleotides exist, valuable insights into possible conformational changes were
gained and the basis for an understanding of their ATP-dependent activity could be established \cite{dawson_07,oldham_07,ward_07,korkhov_12}.
In addition to static structural data, molecular dynamics (MD) simulations have been performed for the NBD components of various ABC transporters \cite{jones_02,
oloo_04,campbell_05,oloo_06,jones_07,ivetac_07,wen_08,jones_09,newstead_09,damas_11}.
It is well established that binding of ATP causes closing of the NBDs and opening of them is obtained after ATP hydrolysis. Although MD simulations
are capable of describing ATP-induced structural changes in the isolated NBD core of ABC transporters, the entire conformational dynamics underlying
the transport cycle involves global motions on long timescales which, due to their heavy demand on computational resources, are still beyond the reach
of such methods \cite{oloo_06review,shaikh_13}.

Therefore, despite extensive experimental and modeling studies, important functional aspects of the operation of ABC transporters are still lacking to date. 
This particularly refers to the coupling of ATP-induced activity of the NBDs to the transport activity inside the TMDs.
Resolving the conformational motions underlying these processes is essential to understand the mechanism of ABC transporters. In this situation, modeling
studies that are based on present structures of transporter proteins and provide approximate descriptions of their dynamics may play an important role.

To overcome the apparent difficulties associated with MD methods, coarse-grained descriptions of protein dynamics have become more and more
popular within the last decades \cite{tozzini_05,tozzini_09}.
A widespread approach to analyze conformational motions in proteins with feasible computational burden is the elastic network model (ENM)
\cite{tirion_96,bahar_97}. In this approximate description, entire amino acids are typically modeled as single point-like particles
and the complex intramolecular interactions are replaced by empirical harmonic potentials which mediate effective interactions between them.
Thus, within the framework of this mechanical model, a protein is viewed as a network of beads connected via deformable elastic springs.
It has been shown that ENMs are able to reproduce well the pattern of residue displacement in protein structures due to thermal fluctuations
and describe ligand-induced collective dynamics in molecular machines \cite{haliloglu_97,tama_01,zheng_03}.
In most former studies, predictions of functional conformational motions in proteins were based on the computation of normal modes of
elastic networks, an approximation which allows for an efficient numerical implementation, but nonetheless may also be prone to fail as reported
earlier \cite{yang_07, togashi_10}. Elastic network normal mode analysis has been previously applied to study nucleotide-dependent conformational
dynamics of the vitamin B$_{12}$ ABC transporter BtuCD \cite{sonne_07}.
More recently, relaxational elastic network models which take into account the full nonlinear network dynamics have been applied to study
mechanochemical motions in protein machines \cite{togashi_07,flechsig_11,duettmann_12}. In a recent study we have used such models
to follow, for the first time, entire operation cycles of an important motor protein in a structurally resolved manner \cite{flechsig_10}.

Here we report investigations performed for the bacterial heme transporter HmuUV that were based on the
relaxational elastic network model. HmuUV is an ABC transporter that is expressed in bacteria as part of their survival strategy since it
presents an efficient agent for the acquisition or iron which typically has a low intracellular concentration but is needed for various chemical
processes \cite{braun_11}.
Based on the crystal structure of the nucleotide-free state of this transporter, which has been recently reported \cite{woo_12}, we have
constructed a ligand-elastic-network model that allowed us to implement interactions with ATP molecules in this protein in a simplified
fashion. The focus in the performed computer experiments was on the investigation of conformational motions induced by binding of ATP ligands
and the dynamical response generated in the transmembrane region, particularly inside the heme translocation cavity and in the
vicinity of the gate to the cytoplasm. Our approach enabled us to trace the formation of an approximate {\it in silico}-version of the HmuUV-2ATP
complex in a structurally resolved way and allowed us to identify effects of intramolecular communication that may explain mechanical aspects of
the gating activity in this ABC transporter.

In particular we find that: i) binding of nucleotides induced generic large-scale conformational motions that were robust, i.e. they were not
sensitive to the specific interaction pattern applied between ligands and the protein within our description, ii) the nucleotide-binding
domains underwent a transition from an open-shape configuration to a closed-shape form while at the same time rotational and tilting
motions of the transmembrane domains were induced, and iii) a unique response inside the central heme translocation cavity, consisting
in channel broadening and opening motions near the cytosolic gate, was generated.

\begin{figure}[t!]
\centering{\includegraphics[width=8.3cm]{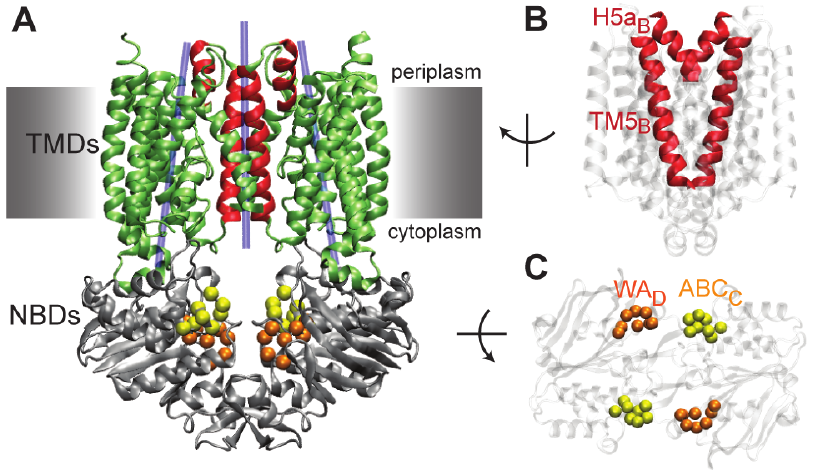}}
\caption{{\bf Nucleotide-free conformation of HmuUV.} 
(A) Structure of HmuUV in ribbon representation. Transmembrane domains are shown in green and nucleotide-binding domains in grey.
The long principal axis of each TMD and the central channel axis are indicated as blue lines. Transmembrane helix 5 and helix 5a of both TMD
chains are highlighted in red. In B) these elements are shown in the side view perspective, indicating that in the nucleotide-free state of the transporter
the gate to the cytoplasm is narrow. (C) The NBDs viewed from the extracellular side. The spatially separated Walker A (WA) and ABC signature
are displayed as golden and yellow beads (only the C$\alpha$ atoms are shown). Two open ATP-binding pockets are formed -- each one by residues
from the Walker A and the ABC motif from different domains.}
\end{figure}

\section*{Methods}

\subsection*{Network construction and dynamics}
For our analysis we have used the crystal structure of the nucleotide-free conformation of HmuUV from {\it Yirsinia pestis} (PDB ID 4G1U).
The coarse-grained protein network was obtained by first replacing each amino acid residue in that structure by a single bead which was
placed at the position of the $\alpha$-carbon atom of the respective residue. These equilibrium positions were denoted by
$\vec{R}_{i}^{(0)}$ for bead $i$. Then, to determine the pattern of network connections, the distance $d_{ij}^{(0)}=|\vec{R}_{i}^{(0)}-\vec{R}_{j}^{(0)}|$
between equilibrium positions of any two beads $i$ and $j$ was compared with a prescribed interaction radius $r_{int}$. If this distance was
below $r_{int}$, the two beads were connected by a deformable link. The network connectivity was stored in matrix ${\bf A}$ with entries
$A_{ij}=1$, if beads $i$ and $j$ are connected, and $A_{ij}=0$, else. Using an interaction radius of $9$\AA, the constructed
network had $1108$ beads and $8029$ deformable links. The majority of links was modeled by elastic springs representing
unbreakable connections. However, for a small subset of links we have made an exception. We have noticed that $11$ links
connect the two TM5 helices at the bottom of the heme translocation pathway close to the cytoplasmic gate. Since in this region, we
did not want to prevent motions which may presumably be important for gating, we have modeled this minor subset
of links by breakable connections. The physical interactions between two beads $i$ and $j$ coupled by an elastic spring were
described by the harmonic potential $U_{ij}^{el}=\frac{\kappa}{2}(d_{ij}-d_{ij}^{(0)})^2$, whereas for the breakable links we have
chosen the Morse potential $U_{ij}^{M}=D\{1-\exp[-a(d_{ij}-d_{ij}^{(0)})]\}^2$, which becomes flat for large enough link deformations.
In these equations, $\kappa$ is the elastic stiffness constant assumed to be equal for all spring connections, $d_{ij}=|\vec{R}_{i}-\vec{R}_{j}|$
is the actual length of the link between beads $i$ and $j$, and parameters $D$ and $a$ determine the shape of the Morse
potential. The dynamics of the protein network is described by a set of Newton's equations of motion considered in the over-damped limit.
They state that the velocity of each network bead is proportional to the sum of forces applied to it. For bead $i$ the differential equation reads as
\begin{equation}
\gamma\frac{d}{dt}\vec{R}_{i}=\sum_{j=1}^{N}A_{ij}\cdot\vec{F}_{ij}+\sigma_{i}\cdot\vec{F}_{i,lig}.
\end{equation}
The first part are contributions from network springs that are connected to bead $i$, with the forces
\begin{eqnarray*}
\vec{F}_{ij}&=&-\frac{\partial}{\partial\vec{R}_{i}}U_{ij}\\
&=&\left\{\begin{array}{l@{\quad\quad}l}
-\kappa\frac{d_{ij}-d_{ij}^{(0)}}{d_{ij}}(\vec{R}_{i}-\vec{R_{j}}) & \text{if } U_{ij}=U_{ij}^{el}\\
-2aD(1-\exp[-a(d_{ij}-d_{ij}^{(0)})])\cdot\\
\exp[-a(d_{ij}-d_{ij}^{(0)})]\frac{\vec{R}_{i}-\vec{R_{j}}}{d_{ij}} & \text{if } U_{ij}=U_{ij}^{M}
.\end{array}\right.
\end{eqnarray*}
The second term $\vec{F}_{i,lig}$ in equ. (1) is the force exerted by an ATP ligand (see next section); the parameter $\sigma_{i}$ equals $1$ if a ligand
is connected to bead $i$, and is $0$ otherwise. In equ. (1), $\gamma$ is the friction coefficient assumed to be equal for all beads and $N$ is the
number of beads in the network ($N=1108$).

\subsection*{Modeling of nucleotide binding}

We established a ligand-elastic-network model of HmuUV, in which an ATP molecule was described by a single ligand bead and its binding process was
modeled by placing it into the center of the binding pocket and connecting it to pocket beads via elastic springs. In the simulations, ligand bead $1$ was
connected to the Walker A motif beads (Gly$^{44}$-Pro-Asn-Gly-Ala-Gly-Lys-Ser) from one NBD and the ABC motif beads (Leu$^{142}$-Ser-Gly-Gly-Glu-Gln-Gln-Arg)
from the other domain. Accordingly, ligand bead $2$ was connected to the corresponding motif beads from the second pocket.
In the simulations, we have chosen the natural lengths of the ligand-springs smaller than their corresponding initial lengths,
implying that initially they were stretched and upon binding the generated forces between each ligand and the selected
pocket beads were attractive.

To check the robustness of our ligand-model, we have followed the formation of the ligand-complexed structure for different
initial ligand-binding conditions. Starting from the native conformation of the HmuUV network, the two ligand beads were
simultaneously placed in their pockets but the natural lengths of ligand springs have been randomly varied in each
realization (i.e., the strength of initial deformations of ligand springs have been altered). More precisely, the following scheme was
implemented in the simulations.
The natural lengths of the ligand springs have been chosen as $d_{lig_{1},p_{k}^{1}}^{(0)}=d_{lig_{1},p_{k}^{1}}^{ini}/\alpha_{p_{k}^{1}}$
for ligand $1$ and $d_{lig_{2},p_{k}^{2}}^{(0)}=d_{lig_{2},p_{k}^{2}}^{ini}/\alpha_{p_{k}^{2}}$ for ligand $2$.
Here, indices $p_{k}^{1}$ and $p_{k}^{2}$ (with $k=1,...,16$) denote the motif beads from pocket $1$, pocket $2$ respectively.
Furthermore, $d_{lig_{1},p_{k}^{1}}^{ini} (d_{lig_{2},p_{k}^{2}}^{ini})$ was the initial length of the spring that connected ligand $1$ 
(ligand $2$) to the motif bead with index $p_{k}^{1} (p_{k}^{2})$ upon its placement. The parameters $\alpha_{p_{k}^{1}}$ and $\alpha_{p_{k}^{2}}$
were random numbers between $2.0$ and $5.0$, ensuring that natural spring lengths were shorter than their initial lengths.
To impose symmetric ligand interactions, these parameters were the same for both ligands (i.e. $\alpha_{p_{k}^{1}}=\alpha_{p_{k}^{2}}$), 
meaning that the spring connecting ligand $1$ to a particular motif bead and the corresponding spring connection for ligand $2$ had
roughly the same prescribed natural length.
Then, as a normalization condition for all realizations, the natural spring lengths were rescaled, such that the amplitude of
the total force $\vec{F}_{lig}^{ini}$ exerted by each ligand upon binding takes a fixed common value, i.e.
$|\vec{F}_{lig_{1}}^{ini}|/\kappa=\sum_{k}(d_{lig_{1},p_{k}^{1}}^{ini}-\tilde{d}_{lig_{1},p_{k}^{1}}^{(0)})=F_{1}$,
$|\vec{F}_{lig_{2}}^{ini}|/\kappa=\sum_{k}(d_{lig_{2},p_{k}^{2}}^{ini}-\tilde{d}_{lig_{2},p_{k}^{2}}^{(0)})=F_{2}$, respectively.
Here, $\tilde{d}_{lig_{1},p_{k}^{1}}^{(0)}$ and $\tilde{d}_{lig_{2},p_{k}^{2}}^{(0)}$
are the rescaled natural lengths of ligand springs and values $F_{1}=0.5\cdot\sum_{k}d_{lig_{1},p_{k}^{1}}^{ini}$ and
$F_{2}=0.5\cdot\sum_{k}d_{lig_{2},p_{k}^{2}}^{ini}$ were used for the forces.
These values correspond to a force that would be generated if each ligand spring in a pocket would be initially stretched by
half the average value of initial spring lengths.
The equation of motion for ligand $1$ was
\begin{equation}
\gamma\frac{d}{dt}\vec{R}_{lig_{1}}=\sum_{k=1}^{16}\vec{F}_{lig_{1},p_{k}^{1}}
\end{equation}
with the elastic forces
\begin{equation*}
\vec{F}_{lig_{1},p_{k}^{1}}=
-\kappa\frac{d_{lig_{1},p_{k}^{1}}-\tilde{d}_{lig_{1},p_{k}^{1}}^{(0)}}{d_{lig_{1},p_{k}^{1}}}(\vec{R}_{lig_{1}}-\vec{R}_{p_{k}^{1}}).
\end{equation*}
The equation for ligand $2$ was the same with index $2$ instead of $1$.
Here, $\vec{R}_{lig_{1}}$ is the actual position of ligand $1$ and $d_{lig_{1},p_{k}^{1}}=|\vec{R}_{lig_{1}}-\vec{R}_{p_{k}^{1}}|$ are the
actual springs lengths. For the ligand beads the same friction coefficient $\gamma$ as for other network beads was assumed.
The additional springs also had the same stiffness constant $\kappa$ as all other elastic springs. According to Newton's `actio and reactio'
principle the forces exerted by the ligand in equ. (1) are related to the forces in equ. (2) by $\vec{F}_{i,lig}=-\vec{F}_{lig,i}$.

Subsequent to binding of the two ligand beads, we have followed the formation of the {\it in silico} HmuUV-2ATP complex by integrating
the set of equations (1) and (2) until a stationary structure of
the network-ligand complex was reached (time moment $t=2500$ in the simulations). This procedure was carried out $50$ times,
each time with a different set of coefficients $\alpha_{p_{k}^{1}}$.
In the simulations, we have considered a rescaling of time ($t'=(\kappa/\gamma)t$) to remove dependencies of the stiffness constant $\kappa$
and the friction coefficient $\gamma$ in equ's (1) and (2). For the numerical integration we have employed an Euler scheme with a time-step of $0.1$.
A numerical value of $D/\kappa=0.05$ was used, and, to ensure that the the Morse potential describing breakable links is harmonic around its
minimum, we have imposed the condition $a=\sqrt{\frac{1}{2D}}$ for parameter $a$.

For the single simulation of the 2ATP-complex formation that was used to illustrate conformational motions upon ligand binding (see Fig. 3)
we provide the ratio $d^{ini}/\tilde{d}^{(0)}$ of initial lengths and rescaled natural lengths of ligand springs in the following table.\\

\begin{tabular}{cc|cc}
Walker A & $l/\tilde{l}^{(0)}$ & ABC & $l/\tilde{l}^{(0)}$\\
\hline
Gly$^{44}$ & $1.92$ & Leu$^{142}$ & $2.23$\\
Pro$^{45}$ & $1.74$ & Ser$^{143}$ & $1.65$\\
Asn$^{46}$ & $2.29$ & Gly$^{144}$ & $1.69$\\
Gly$^{47}$ & $2.28$ & Gly$^{145}$ & $1.60$\\
Ala$^{48}$ & $1.77$ & Glu$^{146}$ & $2.17$\\
Gly$^{49}$ & $2.09$ & Gln$^{147}$ & $1.97$\\
Lys$^{50}$ & $2.29$ & Gln$^{148}$ & $2.34$\\
Ser$^{51}$ & $2.30$ & Arg$^{149}$ & $1.84$
\end{tabular}

\subsection*{Characterization of conformation motions}

To detect global relative motions inside the HmuUV-network, we have measured the distance $d_{NBDs}$ between the centers of mass
of the NBDs. Additionally, to quantify changes in the orientation between NBDs and TMDs, we have determined the angle between
vectors $\vec{v}_{NBDs}$ and $\vec{v}_{TMDs}$ that connect the geometric centers of the NBDs, TMDs, respectively, i.e,
$\varphi_{TMDs,NBDs}=\arccos[(\vec{v}_{TMDs}\cdot\vec{v}_{NBDs})/(|\vec{v}_{TMDs}|\cdot|\vec{v}_{NBDs}|)]$.
To further describe global motions of the TMDs, we have calculated during the simulations the principal axes of each TMD as the
eigenvectors of its inertia matrix.
Motions localized at the central interface of the TMDs, where the heme translocation cavity is located, were monitored by measuring
changes in the orientation (given by a vector connecting residues Ala$^{146}$ and Ile$^{169}$) of the TM5 helices. Additionally, we
have determined the channel profile of the heme cavity in the $50$ different ligand-complexed structures. For that purpose, we have
placed a central axis (defined by the center of mass of the NBDs and that of the TMDs) into the structure and measured the shortest
distance of beads from TMD $1$ (chain A) and beads of TMD $2$ (chain B) to discrete points along this axis. These distances can
be roughly viewed as a pore radius which was plotted as a function of the position along the central axis.

To depict conformations of HmuUV obtained from the simulations in ribbon representation, a structural reconstruction from the
alpha-carbon trace was performed \cite{maupetit_06}. For the visualization the software VMD was used \cite{humphrey_96}.

\section*{Results}

Fig. 1 shows the architecture of the HmuUV heme transporter in its nucleotide-free conformation.
The structure consists of a two-chain cytosolic portion that functions as nucleotide-binding domains (NBDs, chain A and B) and
a pair of chains that correspond to the transmembrane domains (TMDs, chain C and D). In each NBD the conserved Walker A and ABC
signature motifs, which are crucially involved in binding of ATP molecules and their hydrolysis reaction, could be identified. 
In the nucleotide-free conformation of the transporter, these residue motifs are spatially separated and form two open binding
pockets for ATP ligands at the interface of the NBDs (see Fig. 1C). At the interface of the two TMDs, transmembrane helices 5 (TM5)
and helices 5a (H5a) form a cone-shaped cavity which has been assigned the putative heme translocation pathway \cite{woo_12}.
Their outward-facing arrangement indicates that in the nucleotide-free state of the transporter the pathway is open to the periplasm,
whereas it is sealed to the cytoplasm by a narrow gate and heme passage would thus be sterically hindered in this region (Fig. 1B).

We have constructed an elastic network of the bacterial heme transporter HmuUV in its nucleotide-free conformation.
The network consisted of identical beads which represented entire amino acid residues and elastic springs connecting any two beads
if their spatial distance was below a prescribed interaction radius. Details of the network construction can be found in the Methods
section. The elastic network model is an approximative mechanical description of protein dynamics, capable of computationally
handling large proteins with several hundreds (or even thousands) of amino acids and following slow conformational motions inside
them in a structurally resolved way.
The internal motions of proteins in solution are characterized by low Reynolds numbers and therefore dominated by friction.
In fact, it has been shown that inertial effects are negligible on time scales of tens of picoseconds and slower conformational
motions become relaxational \cite{kitao_91}. Therefore, the slow conformational motions with characteristic timescales of milliseconds
taking place in protein machines should be purely damped. Neglecting hydrodynamical effects and thermal fluctuations, the dynamics
of the protein network can therefore be described by mechanical Newton's equations of motion considered in the over-damped limit.
Thus, within the approximate nature of the model, internal motions in proteins are described as processes of conformational relaxation
of its corresponding elastic network.

The aim of this study was to investigate nucleotide-induced conformational motions in the HmuUV ABC transporter, analyze changes inside the
transmembrane structure, and explore the dynamical impact on gating inside the heme translocation pathway. For that purpose, we have
constructed a ligand-elastic-network model of HmuUV that incorporated binding of two ATP mimicking ligands to the pockets inside the
NBDs. This approach will be described in the following section.

\begin{figure}[t!]
\centering{\includegraphics[width=8.3cm]{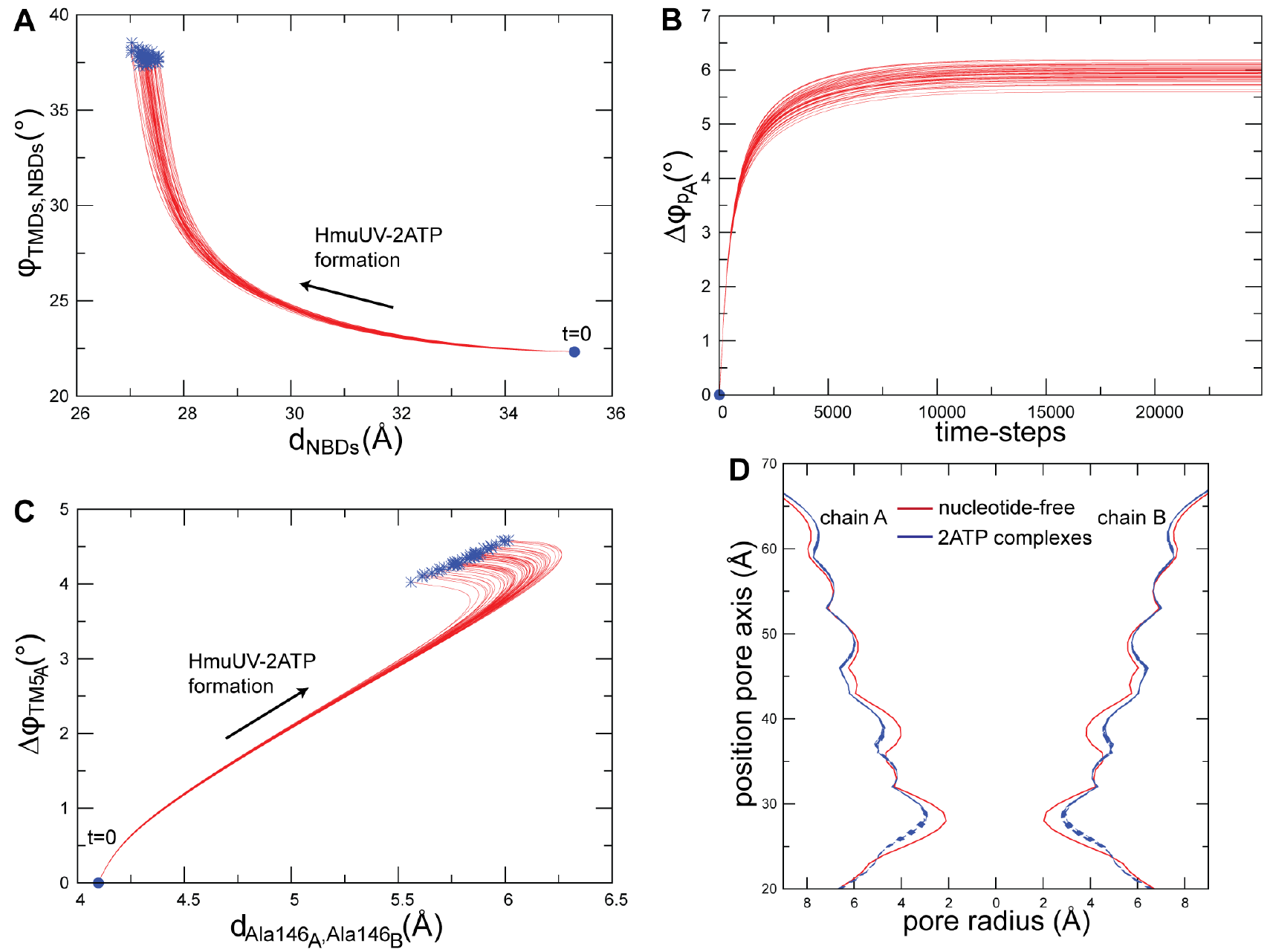}}
\caption{{\bf Nucleotide-induced conformational motions.}
(A-C) Trajectories characterizing conformational motions along the formation of 2ATP-complexes are shown for 50 different ligand-binding realizations.
(A) Trajectories showing changes in the distance $d_{NBDs}$ between the centers of the NBDs, and in the orientation $\varphi_{TMDs,NBDs}$ of the
TMDs relative to the NBDs, subsequent to binding of the two ligand beads. They indicate closing motions of the NBDs and global rotation of the TMDs.
In (B) changes $\Delta\varphi_{p_{A}}$ in the orientation of the long principal axis of one TMD (chain A), reflecting domain tilting motion, are shown as a
function of time. In (C) and (D) changes localized in the heme translocation pathway are displayed. (C) Trajectories showing the distance $d_{Ala146_{A},Ala146_{B}}$
between the Ala$^{146}$ residue of chain A and that of chain B, which are located close to the gate to the cytoplasm, and the change $\Delta\varphi_{TM5_{A}}$
in the orientation of the TM5 helix (chain A), as they evolve after ligand binding. They suggest gate opening as a result of tilting of the TM5 helices. In (D) the
channel profiles of the 50 different 2ATP-complexes (blue lines) are compared with that of the ligand-free structure (red lines).}
\end{figure}

\subsection*{ATP-ligand binding and related conformational motions}

The HmuUV protein could not yet be co-crystallized with ATP-analogs. Therefore, interactions with ATP and the effect of binding
and its hydrolysis for this ABC transporter are hitherto unknown. However, the two nucleotide-binding pockets are well-defined
by the presence of the conserved Walker A and ABC signature motifs located in each NBD. Moreover, it has been shown previously,
that most ABC transporters bind two ATP molecules in each single catalytic cycle \cite{patzlaff_03}.
On the other side, the coarse-grained nature of the elastic network model does not allow us to resolve the complicated interactions
between the protein and ATP molecules in full detail. To investigate conformational motions induced by binding of ATP molecules,
we have constructed a dynamical ligand-elastic-network model of HmuUV that incorporated binding of two ATP nucleotides to the NBDs
in a simplified fashion. Details of the modeling and the network dynamics are provided in the Methods section and the limitations of our
approach are mentioned in the Discussion section.

\begin{figure*}[t!]
\centering{\includegraphics[width=17.35cm]{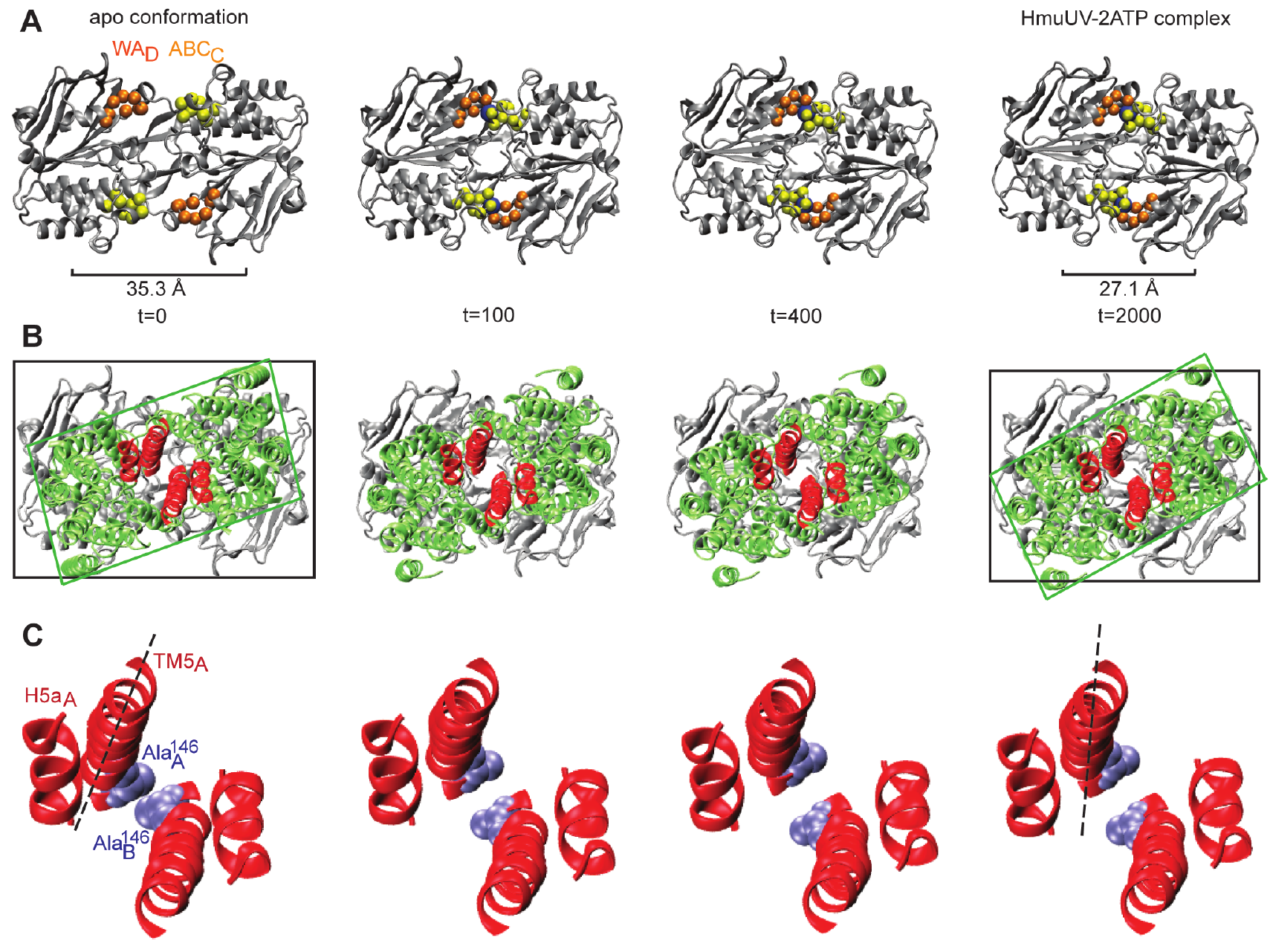}}
\caption{{\bf Characterization of HmuUV-2ATP complex.} Snapshots from a single simulation showing conformational motions along the formation of the
HmuUV-2ATP complex are depicted in the extracellular perspective. In all panels the left picture corresponds to the ligand-free conformation of HmuUV.
(A) Switching of the NBDs from the open-dimer to the tight-dimer form along complex formation. The Walker A and ABC signature motifs of the ATP binding
pockets are shown as orange and yellow beads (only the $C{\alpha}$ atoms are shown) and the two ATP ligands are shown as blue beads. In (B) changes
in the orientation of the TMDs (shown in green) with respect to the NBDs (shown in grey), as indicated by rectangles, are best seen. In (C) the TM5 and H5a
helices of the heme pathway (shown in red color in panel (B)) are displayed. Additionally, the Ala$^{146}$ residues which mark the entry gate to the
cytoplasm are shown. In this perspective gate-opening due to tilting motions of the TM5 helices becomes apparent.}
\end{figure*}

Similar to our previous publication, where entire operation cycles of the hepatitis C virus helicase motor protein were investigated using
ENM modeling \cite{flechsig_10}, binding of ATP was phenomenologically accounted for in the present study. Consistent with the coarse-grained
character of the elastic network model, we have described an ATP-ligand by a single bead and modeled its binding process to the ATP-pocket
by allowing it to physically interact with network beads present there.  We have selected those beads as interaction spots that corresponded to
residues from the conserved sequence motifs mentioned above. More precisely, a ligand could interact with beads of the Walker A motif from
one NBD and beads of the ABC motif from the other NBD. Binding of the two ATP-ligands was implemented by simultaneously placing each ligand
bead into its binding pocket and creating spring connections to the motif beads. These additional elastic springs were initially stretched, so that
upon binding attractive forces between each ligand and the selected pocket beads were generated. Therefore, as a result of the binding process,
network deformations were initiated. First localized in the two binding pockets, they gradually spread and conformational relaxation motions of the
entire protein network took place until a stationary structure of the network-2ATP complex was reached.

To follow the formation of an approximate {\it in silico}-version of the HmuUV-2ATP complex upon binding of the two ligand beads was the focus of our computer experiments.
We have numerically integrated the set of equations of motion to obtain the position of all network beads at every instant of time.
Since in our approach interactions between ATP ligands and the pocket were modeled only roughly, we have tested the robustness
of the ligand network description and its predicted conformational changes by following the formation of the ligand-bound complex
several times, each time starting from the nucleotide-free network. In each realization different ligand-binding patterns were applied
by randomly varying initial deformations of the ligand springs (see Methods).
To characterize conformational motions accompanying binding of the two ligand-beads we have traced several geometric variables
during the simulations, reflecting both global domain rearrangements as well as structural changes localized in the heme translocation
cavity of the TMDs (see Methods). A process of conformational relaxation corresponding to one realization of the HmuUV-2ATP complex formation
could therefore be visualized by a single trajectory in the space of these variables. The results of our simulations are shown in Fig. 2.

For all different realizations of ATP-ligand binding conditions we observed similar dynamical behavior of the transporter along the
formation of the ligand-complexed structures, as evidenced by the low dispersion in the trajectories that monitor changes of the protein
structure. Subsequent to binding of the two ATP-beads, large-amplitude relative motions of the
two NBDs were induced with the distance between their centers of mass decreasing about $8$\AA\ (see Fig. 2A). At the same time,
the angle between the TMDs and the NBDs uniquely changed, indicating global rotational motions of the TMDs relative to the NBDs.
We find that latter motions are first small after ligand binding but become pronounced when the steady conformations of the
complexes are approached. In addition to the rotational dynamics, we observed tilting motions of each individual TMD as evidenced
by a change in the orientation of the long principal axes of $\sim 6^{\circ}$ (see Fig. 2B, shown only for one TMD chain).

These relative motions of the TMDs also imply structural changes localized inside the central heme pathway. In the simulations, we have detected
systematic changes in the orientation of the TM5 helices, while at the same time, the distance between the Ala$^{146}$ residues increased
about $2$\AA\ (see Fig. 2C). The latter residues are located at the bottom of the heme cavity and constitute the gate to the cytoplasm.
To display changes inside the heme cavity conveniently, we have plotted the channel profile for the obtained ligand-complexed structures
together with that of the native conformation of the transporter in Figure 2D. While changes in the upper channel region close to the
periplasm were marginal, a clear broadening of the middle part of the cavity was observed in the conformations that corresponded
to the liganded complexes. Most notable was the opening of the channel close to the cytoplasmic region.

\subsection*{Characterization of ligand-bound complex}

To illustrate conformational motions in the course of ligand binding we provide structural snapshots from a single simulation of the 2ATP-complex
formation (i.e. for one particular ligand binding configuration, see Methods section). The structure of the transporter is shown in Fig. 3
in three views at different time moments along the complex formation.

When comparing the nucleotide-free structure of the transporter with that of the ligand-bound complex, the most notable change
is observed in the conformation of the NBDs which differ by a RMSD value of $5.6$\AA\ (see Fig. 3A). In the native conformation,
the two domains are well-separated by a distance of $35.3$\AA\ resembling an open-shape configuration. The ATP-binding pockets
are open and have gyration radii $R_{g}^{open}\sim10$\AA. In the ligand-bound complex, in contrast, the NBDs are in close contact having
a distance of $27.1$\AA\ between their centers and forming a closed-shape structure. The binding pockets in this conformation
are closed ($R_{g}^{closed}\sim6$\AA) and the ligand-beads are sandwiched between the Walker A and ABC motifs.
In existing structures of ABC transporters complexed with nucleotides, the NBDs are also found in a closed conformation. In the
AMP-PNP bound vitamin B$_{12}$ transporter BtuCD-F, which is homolog to HmuUV, the NBDs are separated by a distance of
$27.9$\AA\ ($R_{g}^{closed}\sim6.8$\AA). From the structures of the bacterial lipid flippase MsbA-AMPPNP and the multidrug transporter
Sav1866-ADP, we extract corresponding distances of $27.7$\AA\ ($R_{g}^{closed}\sim6.9$\AA) and $26.4$\AA\ ($R_{g}^{closed}\sim6.7$\AA),
respectively.

\begin{figure}[t!]
\centering{\includegraphics[width=8.3cm]{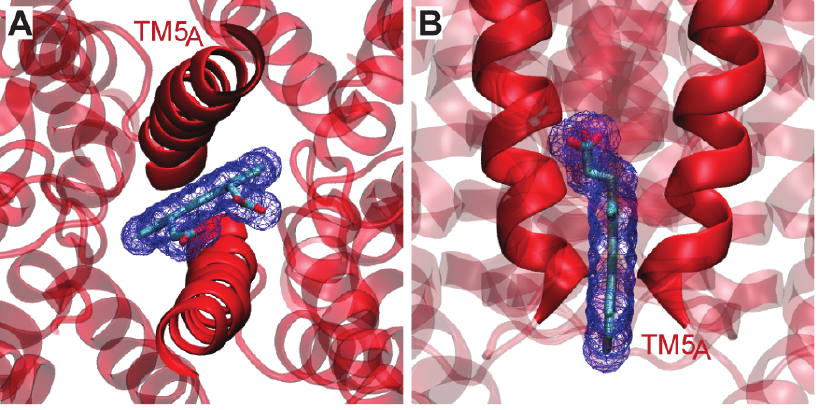}}
\caption{{\bf Heme translocation.} Close-up view of the heme cavity with a single heme substrate modeled into the bottom region at the cytoplasmic gate.
In (A) a perspective similar to that in Fig. 3C was chosen. The TM5 helices are shown as solid ribbons whereas the rest of the TMDs is transparent. The
heme structure is shown as sticks together with its wireframe surface representation (blue). In (B) a side view perspective similar to that in Fig. 1B was
employed to show the alignment of the heme molecule near the cytoplasmic gate.}
\end{figure}

Further structural comparison in HmuUV shows also a remarkable change in the orientation of the TMDs relative to the NBDs resulting
from ligand-induced rotational motion (Fig. 3B). The TMDs in the nucleotide-free conformation and the ligand-bound complex compare by
a RMSD value of $2.9$\AA. Regarding the heme translocation cavity we note that the position of H5a helices remain almost unchanged,
whereas a significant change in the orientation of the TM5 helices is present (Fig. 3C). As a result of tilting motions of these elements,
the distance between the Ala$^{146}$ residues located at the bottom of the cavity changed from $4.1$\AA\ to $6.0$\AA. Therefore,
the cytosolic gate which was closed in the ligand-free structure is found open in the ligand-bound complex. The implication for heme
passage through the cavity is reported in the next section.

\subsection*{Implications for heme passage}

It has been previously reported, that a heme molecule can be placed into the broad upper region of the central cavity close to the
extracellular entry without producing steric clashes \cite{woo_12}.
In our simulations, as a result of ligand binding to the HmuuV structure, we have identified considerable motions inside the heme
translocation cavity associated with an opening of the gate to the cytoplasm. To check whether these structural changes may be
of functional importance for the passage of heme substrates, we have modeled a single heme molecule into the structure. Being
properly aligned, we found that the heme structure readily fitted into the space between the TM5 helices at the gate to the cytoplasm
(see Fig. 4). This indicates that the size of the observed tilting motions of TM5 helices inside the cavity is sufficient to explain mechanical
aspects of heme translocation in the HmuuV transporter: In the nucleotide-free conformation of HmuUV, the gate to the cytoplasm is
closed preventing heme substrates to enter the cell. In the ligand-bound complex, in contrast, the gate is open making heme passage
possible.

\section*{Discussion}

In the present study we have used coarse-grained molecular dynamics simulations to investigate conformational dynamics of
the bacterial ABC transporter HmuUV from {\it Yirsinia pestis} that organizes the uptake of heme substrates into the cytoplasm.
The crystal structure of HmuUV in its nucleotide-free conformation has been recently reported and was used by us to establish
a dynamical description for this protein machine. Based on the elastic-network model, in which a protein is viewed as a meshwork
of beads connected by deformable springs, we have implemented an extension that allowed us to take into account interactions
between the HmuUV transporter and ATP ligands in an approximate fashion. We were then able to follow the formation of the
HmuUV-2ATP complex {\it in silico} and study nucleotide-induced dynamics underlying the activity in this ABC transporter. It is
remarkable that despite the gross simplification present in our approach, we were able to identify functional conformational
motions.

We found that binding of ligands induced large-amplitude structural rearrangements taking place inside the protein. The most
significant dynamical response was detected inside the two nucleotide-binding domains, which subsequent to ligand binding, underwent
substantial motions bringing them from an open-shaped conformation in which they were well separated, to a closed-shape form
where they were found in tight contact. At the same time, the orientation of the transmembrane domains relative to the
nucleotide-binding domains changed as a result of rotational and tilting motions. In addition to these global structural changes, we have
detected motions localized at the interface between the two transmembrane domains and inside the heme cavity, which may be
of functional importance for the translocation of heme substrates. There, tilting of the TM5 helices lead to channel broadening and
resulted in opening of the gate to the cytoplasm, which was closed in nucleotide-free conformation and thus sterically hindered heme
passage.

The identified conformational motions were robust and not sensitive to the specific interaction pattern that was applied between the
ligands and HmuUV within our ligand-network description. This indicates that interactions with nucleotides in this membrane
transporter generate a generic response that consists in similar well-organized and ordered domain motions. This property may assure
that the transporter is capable of maintaining its cyclic operation despite the presence of possible perturbations.

\subsection*{Possible transport scenario}

Based on our findings we propose that the HmuUV transporter behaves like a `simple' mechanical device in which, induced by binding
of ATP ligands, linear motions of the nucleotide-binding domains are translated into rotational motions and internal tilting dynamics of the
transmembrane domains. The structural changes have functional consequences: In the nucleotide-free conformation of HmuUV, heme
passage is prevented since the pathway to the cytoplasm is sealed by a narrow gate. Although ATP molecules are processed inside the
nucleotide-binding domains, the induced conformational changes are robustly communicated through the protein structure to the remote
heme pathway where they trigger gate opening by a mechanism that involves tilting motions of essential TM5 helices.
Thus in the ATP-complexed structure heme substrates would be able to pass the barrier and enter the cytoplasm. The nucleotide-binding
domains are in close contact in this conformation, with the nucleotides being sandwiched between the Walker A and ABC signature motifs
and ready for their hydrolysis. Once the hydrolysis reaction has occurred and the chemical products become dissociated, the transporter
would be able to return to its initial conformation, closing the cytoplasmic gate and restoring the open-shape configuration of the nucleotide-binding
domains. These domains are then again ready to bind other ATP molecules and initiate the next operation cycle of the transporter.

Comparing the results from our computer experiments with experimental data acquired for other ABC proteins, we firstly note that
the switching of the nucleotide-binding domains from an open to a tight dimer conformation upon binding of nucleotides, as expected from
the analysis of available crystal structures, is reproduced also in the case of HmuUV. In particular, the arrangement of the nucleotide-binding domains in the
vitamin B$_{12}$ transporter BtuCD-F-AMPPNP, the bacterial lipid flippase MsbA-AMPPNP and the multidrug transporter Sav1866-ADP closely
resembles that of the HmuUV-2ATP complex, as predicted by our simulations.

While the nucleotide-induced approaching of the NBDs represents a well-established aspect common for the operation of all ABC transporters,
the coupling between motions of the NBDs and the TMDs amongst them is understood less well and, hence, the mechanism underlying their
transport cycle is still controversially discussed. Two opposing models have been proposed based on crystal structures of free and nucleotide-complexed
conformations of transporters. The MalK transport model, deduced from the analysis of the maltose transporter system \cite{chen_01,chen_03},
predicts that ATP binding to the NBDs, and their subsequent transition from the open to a tight dimer form, triggers closing of the TMDs at the
cytoplasm interface and opening towards the periplasmic side. The BtuCD model, introduced by Locher et al. based on data for the vitamin B$_{12}$ ABC
transporter \cite{locher_02}, on the other side, claims the opposite response in the TMDs upon ATP binding to the NBDs - opening of the gate to
cytoplasm and closing at the periplasm region.
Significant changes in the TMDs near the periplasmic side, as suggested by those models, could not be detected in our simulations for HmuUV. Apart
from that, our findings are consistent with the BtuCD model whereas the disagree with the MalK mechanism. According to the above described possible
transport scenario for HmuUV, heme transport would be controlled by switching between ligand-free and ATP-complexed states and the concomitant 
cytoplasmic TMD gating by a valve-like mechanism. For this process, changes of the TM5 helices are crucial. The role of the transmembrane helices for
substrate transport has been previously debated and their mobility seems to be essential for gating in ABC transporters \cite{dawson_06,ward_07,korkhov_12}.
To check whether tilting motions of the TM5 helices are the major cause of gating in HmuUV and to further elucidate details of heme transport,
it would certainly be important to determine the structure of this transporter with ATP-analogs. Furthermore, our predictions of dynamical changes in HmuUV
can possibly be checked in site-directed spin labeling electron paramagnetic resonance (EPR) experiments.

\subsection*{Model limitations and concluding remarks}

Any coarse-grained model is obtained through approximations and therefore underlies limitations. The elastic-network model provides a
mechanical description of nucleotide-induced dynamics in protein machines whereas, as the price paid for of the applied simplifications, the
involved chemical processes cannot be resolved.
In our study the elastic-network model could not resolve the chemical details that are important to correctly model interactions of HmuUV with ATP
molecules. Therefore, similar to our previous elastic-network study where full operation cycles of the hepatitis C virus helicase protein motor could
be modeled \cite{flechsig_10}, the process of ATP-binding was emulated only roughly and physical interactions were accounted for by empiric effective potentials.
On the other hand our results suggest that details of the ligand-protein interactions are not particularly important when considering nucleotide-induced
global conformational dynamics; the same robust and generic motions were generated, irrespectively of the applied interaction pattern. This observation
agrees with our elastic-network studies of mechano-chemical dynamics of other molecular machines \cite{flechsig_10,flechsig_11}.
It is indeed remarkable that a purely mechanical description can still capture essential properties of the functional dynamics in proteins.

Generally, our study demonstrates the feasibility of coarse-grained mechanical models in computer experiments of molecular machines.
Apparently they cannot replace all-atom simulations. However, given the situation that detailed molecular dynamics methods are not yet
capable of describing the entire conformational motions which underlie the transport cycle in ABC transporters, coarser models are potentially
useful to explore important aspects of their operation. To study dynamical properties of membrane machines more realistically in computer
experiments, it can be attempted in the future to combine elastic-network models with coarse-grained descriptions of the membrane,
which would additionally allow to investigate their interplay.

\end{document}